\documentstyle[12pt,aaspp4,flushrt]{article}
\lefthead{JENKINS}
\righthead{CORRECTING APPARENT OPTICAL DEPTHS}
\begin{document}
\title{A Procedure for Correcting the Apparent Optical Depths of
Moderately Saturated Interstellar Absorption Lines}
\author{Edward B. Jenkins}
\affil{Princeton University Observatory\\
Princeton, NJ 08544-1001; ebj@astro.princeton.edu}
\begin{abstract}
Presently, most observations of absorption lines from interstellar and
intergalactic matter have sufficient resolution to show most of the
structure at differing radial velocities of the absorber.  This added
information allows one to go beyond the practice of just obtaining
equivalent widths.  As with measurements of $W_\lambda$, however, it is
important to sense and correct for the fact that some parts of a profile
may arise from absorption peaks that are strong enough to be saturated. 
This effect may be unrecognized, or at least underappreciated, in those
cases where the narrowest velocity structures are degraded by the
convolution of the true spectrum by the instrumental profile.

Using a procedure that is virtually identical to the curve of growth
method for equivalent widths, one can compare at any velocity the
apparent optical depths $\tau_a$ of two lines that have significantly
different transition probabilities.  If their ratio is smaller than the
ratio of the lines' values of $f\lambda$, the actual saturation is more
severe than that indicated by the values of $\tau_a$.  This paper
describes a simple procedure for selectively boosting $\tau_a$ of the
weaker of the two lines so that unresolved saturated structure is
accounted for.  This enables one to obtain a very nearly correct answer
for the column density per unit velocity. (The lost velocity detail is
not restored however.)  Two synthetic, test examples of very complex,
saturated profiles are analyzed with this method to show how well it
works.  A demonstration with real observations is also presented.  An
explicit, easily-computed formula that is a very close approximation to
the real correction factors is given, to make data analysis and error
estimation more convenient.
\end{abstract}

\section{Introduction}\label{intro}

Over recent years, improvements in spectrographs and detectors have
brought forth substantial gains in the quality of observations of
absorption features arising from either interstellar gases in front of
stars in our Galaxy or material in very distant systems in front of
quasars.  Most modern observations of these features have good
signal-to-noise ratios, accurate determinations of the zero intensity
level, and sufficient wavelength resolution to break the overall
absorption profiles into subcomponents at different Doppler shifts. 
These advances have allowed us to progress beyond the simple practice of
measuring and interpreting just the total equivalent widths of the
absorptions. Now, with the ability to discern the added dimension of
velocity in the absorption features, an observer is presented with new
opportunities for more detailed interpretations.  With this expansion,
however, come new challenges and responsibilities, ones that extend
beyond the framework of analysis techniques that were connected with
equivalent widths.

Except for features that we are sure must arise from regions with
elevated temperatures, we are rarely confident that all of the
substructures within the radial velocity peaks have been completely
discerned by the spectrograph.  There is evidence that, as a rule,
observations taken at successively higher resolutions reveal finer
details than those registered before; good examples can be seen in the
interstellar Na~I absorption features recorded by Wayte, Wynne-Jones \&
Blades \markcite{2056} (1978), Blades, Wynne-Jones \& Wayte
\markcite{304} (1980), Welty, Hobbs \& Kulkarni \markcite{262} (1994)
and Barlow, et al. \markcite{3122} (1995) and some molecular lines
observed by Crawford, et al. \markcite{2651} (1994) and Crane, Lambert
\& Sheffer \markcite{304} (1995).  Ultimately, the intrinsic dispersion
of Doppler velocities (partly thermal, partly turbulent) may be the only
limiting factor in the fineness of the real features.  With typical
temperatures of cool gas complexes extending below 100K and negligible
turbulence, we can expect velocity dispersion parameters $b$ that could
be as small as 0.2 km~s$^{-1}$ (for atoms with a mass of about 40 amu),
a value that is still significantly narrower than most present-day
instrumental profiles.  Unfortunately, as we shall see in the discussion
that follows, the consequences of instrumental smoothing are more
serious than just a loss of velocity detail.  Observers must often face
the challenge of determining how badly saturated the absorptions were
before the smoothing took place.  This is an important step in deriving
trustworthy conclusions about the amount of material that caused the
absorption.

An observer's ultimate objective is usually to determine not only the
total column density $N$ of an absorber, but also how the atoms, ions or
molecules are distributed over different radial velocities.  For the
original form of the spectrum $I(v)$ that has not been degraded by a
convolution with the instrumental profile, the column density as a
function of Doppler velocity $v=(\lambda - \lambda_0)/(c\lambda_0)$ is
equal to the absorption feature's optical depth $\tau(v)=\ln [I_0/I(v)]$
multiplied by the constant factor $(m_ec)/(\pi e^2f\lambda)$ ($I_0$ is
the intensity of the unabsorbed continuum).  What one observes in actual
practice, however, is an apparent intensity $I_a(v)$ that is a smoothed
form of the real intensity profile $I(v)$.  Even so, as long as the
smoothing is not too severe, one can derive an approximate
representation that is called the {\it apparent} optical depth
$\tau_a(\lambda) = \ln[I_0/I_a(v)]$, an interpretative concept first
used for high resolution recordings of lines in the visible part of the
spectrum by Hobbs \markcite{1080,1081,1083,1085,1086} (1971, 1972, 1973,
1974a, b) and later invoked for UV lines by Savage, et al.
\markcite{1149} (1989), Jenkins, et al. \markcite{1367} (1989), Savage,
Massa \& Sembach \markcite{29} (1990), Joseph \& Jenkins \markcite{1728}
(1991),  Sembach, Savage \& Massa \markcite{119} (1991) and Tripp,
Sembach \& Savage \markcite{2551} (1993) in their analysis of some IUE
and IMAPS data.  More recently, spectra of exceptionally good quality
and resolution have been produced by the GHRS echelle spectrograph on
the Hubble Space Telescope, and representations of $\tau_a$ have been
important tools for understanding these data
\markcite{1965,1966,2369,2519,2630,311} (Cardelli et al. 1991; Savage et
al. 1991; Savage, Cardelli, \& Sofia 1992; Sofia, Savage, \& Cardelli
1993; Savage, Sembach, \& Cardelli 1994; Cardelli \& Savage 1995). 
Important properties of $\tau_a$ have been explained in detail by Savage
\& Sembach \markcite{110} (1991).

The papers cited above have made it clear that apparent optical depths
are useful functions for deriving column densities and extracting linear
representations for all of the kinematical information that is
available.  The information conveyed by the $\tau_a$ functions
represents a significant improvement over the single numbers that
signify the equivalent widths of entire profiles or resolved pieces of
profiles.  Nevertheless, we must be aware of some limitations that arise
in certain circumstances \markcite{1367,1728} (Jenkins et al. 1989;
Joseph \& Jenkins 1991).  The real physical processes that created the
recorded intensities consisted of an exponential attenuation of the
light, followed by an instrumental smearing of the spectrum.  The
derivation of $\tau_a$ is an attempt to reconstruct a linear
representation for the amount of absorbing material by unraveling the
exponential absorption law, but it disregards the convolution by the
instrumental profile that followed.  Normally, we are accustomed to
interpreting functions where there is simply a loss of detail caused by
smoothing.  But unfortunately $\tau_a$ does {\it not} represent just a
smoothed version of the real $\tau$.  Instead, we find that the
smoothing has deaccentuated the extremes in $\tau$, and the nonlinear
operation used to construct $\tau_a$ creates a representation of $\tau$
that is both smoothed {\it and distorted}.  There are only two
circumstances where $\tau_a$ represents an unbiased reflection of the
sought-after distribution: (1) All of the velocity details of the
profile were fully resolved so that $\tau_a$ is identical to $\tau$, or
(2) $\tau\ll 1$ everywhere, so that the $\ln[I_0/I(\lambda)]$ is
essentially equivalent to the linear representation
$[I_0-I(\lambda)]/I_0$ whose integral over any $\lambda$ interval is not
changed by the convolution operation.  In the first case, all of the
information is recovered, while in the second, the only repercussion is
a loss of velocity detail.

In short, in the course of interpreting $\tau_a(v)$ an observer must be
vigilant about the possible loss of evidence that narrow peaks in
absorption are badly saturated.  There is a danger that instrumental
smearing has created a picture where the reduction of intensity {\it
appears} to be weak and thus far from saturation.  The same argument
holds for stronger features.  Even if one can sense that some saturation
must be evident because the intensity is significantly below the
continuum level, smoothing of the bumps could cause one to underestimate
its severity and then misjudge the actual amount of material in the line
of sight.

A straightforward way to sense and measure the amount of hidden
saturation is to observe two or more lines with differing transition
probabilities from the same species.  If, at any velocity $v$, the
apparent optical optical depths in the smoothed spectra exhibit a
scaling that is weaker than the progression of the respective lines'
$f\lambda$ values, then there is good reason to believe that in some
places the unresolved saturated structures are stronger than a general
level suggested by the apparent (smoothed) intensity values.  The object
of this paper is to demonstrate how one can correct for this effect and
derive reasonably accurate representations of column density as a
function of velocity.  The method to be outlined has a relationship with
the conventional curve of growth analysis for equivalent widths that is
stronger than just a simple analogy.  As the arguments in
\S\ref{concepts} will show, the two methods have nearly identical
mathematical foundations.  The principal advantage of correcting
$\tau_a$ rather than $W_\lambda$ is that we do not sacrifice the
information contained in velocity peaks that can be resolved.  Hence one
can explore, for instance, how the abundances of different species
change with velocity, rather than just determining the overall abundance
ratios at all velocities within some large complex of components.

In their instructive overview on how to derive, interpret and exploit
the apparent optical depths of interstellar features, Savage \& Sembach
\markcite{110} (1991) likewise addressed the problem of how to cope with
the misrepresentations of the real optical depth levels caused by
instrumental smearing.  They proposed that one should measure the
disparity in inferred column densities {\it integrated over velocity}
for two lines and then apply, according to a specific prescription, a
global, multiplicative (upward) correction to the entire profile of the
weaker line.\footnote{This correction procedure may be applied to any
two lines of arbitrarily different strengths (within reason), but Savage
\& Sembach supplied correction factors only for lines that had a 2:1
ratio for $f\lambda$.}  The method advocated here uses a different
approach that is an improvement over the one described by Sembach \&
Savage.  Corrections are applied on the spot at individual velocities,
without regard to what is happening elsewhere.  The two methods will be
compared in \S\ref{comparison}.

A very different tactic for analyzing saturated, blended features is to
build a model of the real absorption complex by defining such parameters
as the strengths, widths and velocity centroids of individual components
\markcite{1769,1819,19,1872,2671,1988,2481,262,2462,2701,250,304}
(Vidal-Madjar et al. 1977; Ferlet et al. 1980; Welsh, Vedder, \&
Vallerga 1990; Hobbs \& Welty 1991; Welsh et al. 1991; Welty, Hobbs, \&
York 1991; Spitzer \& Fitzpatrick 1993, 1995; Vallerga et al. 1993;
Fitzpatrick \& Spitzer 1994; Welty, Hobbs, \& Kulkarni 1994; Crane,
Lambert, \& Sheffer 1995).  One then solves (or searches) for a minimum
in the $\chi^2$ values as parameters for the theoretical representation
of the instrumentally blended complex are compared with the
observations.  In cases where independent information can help to
constrain the choice of free parameters (such as much higher resolution
observations of other species), this method can be successful.  While
such model building has the potential of helping us to understand some
details that may not be evident in a display of smoothed optical depths,
it has the disadvantage of usually relying on human judgement to define
the constraints on the parameters and the method of converging to a
minimum $\chi^2$.  Also, the models contain specific assumptions about
the functional forms of the components, with the usual choice being a
Gaussian (or Voigt) profile.  By contrast, the derivation of $\tau_a$ is
a simple, mechanical process that places no such requirements on the
investigator and does not rely on any specific models, even when the
corrections discussed later in this paper are implemented.

\section{Basic Concepts}\label{concepts}

For the purposes of discussion, we shall address the problem of working
with the optical depths of two lines that have values of $f\lambda$ that
differ by a factor of 2.  This situation is frequently encountered in
astronomical spectroscopy, such as when the strong $^2{\rm S}-{}^2{\rm
P}^0$ resonance doublets of lithium-like atoms and ions are observed. 
The arguments apply equally well to other line strength ratios, within
reasonable limits that are set by errors that arise either from noise or
systematic measurement problems.

In \S\ref{single_line} below, we start with a trivial example of how to
analyze a single Gaussian profile.  In sections that follow, we make use
of some simple theorems to address progressively more complex
situations, ending up with arbitrarily strong profiles with very
complicated shapes. 

\subsection{A Single, Saturated Gaussian Profile}\label{single_line} 

Before approaching the problem of correcting optical depths, we should
first review the basic principles of the doublet ratio method
\markcite{3103, 3105, 3104, 3101} (Uns\"old, Struve, \& Elvey 1930;
Beals 1936; Wilson \& Merrill 1937; Str\"omgren 1948), a classical
analysis that is applied to the equivalent widths of the two members of
a doublet.  Strictly speaking, the analysis is correct only for a Voigt
intensity profile created by a Gaussian 1-dimensional velocity
distribution (or two or more such profiles caused by {\it identical},
well separated Gaussian components).  In practical situations where
there is no better choice, it is customary for investigators to make the
implicit assumption that the velocity structure of the absorbing
material is very similar to that of a single Gaussian.  When a line is
optically thin everywhere
\begin{equation}\label{thin}
{W_\lambda\over \lambda}={\pi e^2f\lambda N\over m_ec^2}~.
\end{equation}
However when there is enough material and a low enough velocity
dispersion $b$ to make the lines saturated, any line with a central
optical depth
\begin{equation}
\tau_0={\pi^{1/2}e^2f\lambda \over m_e c}\left({N\over b}\right)
\end{equation}\label{tau0}
should have an equivalent width given by
\begin{equation}\label{thick}
{W_\lambda\over \lambda}={2b F(\tau_0)\over c}
\end{equation}
where
\begin{eqnarray}\label{Ftau0}
F(\tau_0) & = & \int^\infty_0 [ 1 - \exp(-\tau_0 e^{-x^2})]dx
\nonumber\\
& = & {\pi^{1/2}\over 2}\sum^\infty_{n=1}{(-1)^{n-1}\tau_0^n \over
n!n^{1/2}}~.
\end{eqnarray}
From the ratio
\begin{equation}\label{R}
R=F(2\tau_0)/F(\tau_0)
\end{equation}
of the doublet's two equivalent widths, one can calculate a correction
factor
\begin{equation}\label{C}
C_R={\pi^{1/2}\tau_0\over 2F(\tau_0)}
\end{equation}
that would be needed to enhance the weaker line's $W_\lambda$ so that
Eq.~\ref{thin} for low optical depths would apply.  (In this case,
$\tau_0$ is the central optical depth of the weak line and the factor 2
in Eq.~\ref{R} reflects the ratio of the two lines' $f\lambda$ values.)

Now that the concept of doublet ratio corrections has been introduced,
we move on to address how we would operate with the two $\tau_a(v)$
functions instead of just a pair of numbers representing the absorption
over all velocity, i.e., the two $W_\lambda$'s.  Again, we consider an
observation of moderately saturated lines in a (2:1) doublet, where the
column density of the absorber has a velocity distribution that is a
single Gaussian and the feature has a central optical depth $\tau_0$ for
the weaker line.

In the limit where the instrumental profile is very much broader than
the widths of the lines, the two smoothed absorption functions
$[I_0-I_a(v)]/I_0$ have several important properties.  First, they have
shapes that are virtually identical to the instrumental smearing
function and areas equal to the lines' equivalent widths.  Second, by
virtue of the fact that the extreme smearing has degraded the amplitudes
of these functions so that their peak values are very much less than
one, they are close approximations to the $\tau_a(v)$ functions.  It
then follows that at every $v$ the ratios of the two lines'
$\tau_a(v)$'s are identical to the ratio of the equivalent widths.  If
we now take this ratio $R$ and solve Eqs.~\ref{Ftau0} and \ref{R} to
find $\tau_0$, we may derive the correction factor from Eq.~\ref{C} and
evaluate $N$ according to
\begin{equation}\label{N}
N={m_ec\over \pi e^2f\lambda}\int C_R\tau_a(v) dv~,
\end{equation}
where here $C_R\tau_a(v)$ replaces $\tau(v)$ in the usual equation for
deriving $N$ from the integral of a true optical depth over all
velocity.  While this cumbersome procedure for obtaining $N$ is
mathematically equivalent to invoking the doublet ratio method on
equivalent widths, it serves as a trivial but good introductory
illustration of the principle of correcting $\tau_a(v)$.

If we move to the opposite extreme of having a smearing function that is
much narrower than the velocity spread of $\tau(v)$, we find that
$\tau_a(v)$ of the strong line is always twice the value of $\tau_a(v)$
of the weak line because $\tau_a(v)$ is practically identical to the
true $\tau(v)$.  Here, one would obtain $N$ by evaluating Eq.~\ref{N}
with $C_R=1$.

In the two preceding extreme cases of very poor and very good
resolution, $C_R$ does not change with $v$.  In the intermediate case
where the line is only partly degraded by instrumental smearing, both
$R$ and $C_R$ vary. (Henceforth, we will work with a modified notation
$R(v)$ and $C_R(v)$ as a reminder that these quantitites change with
velocity).  Also, it is no longer possible to show by elementary
arguments that a point-by-point correction will indeed make Eq.~\ref{N}
work.  In fact, numerical simulations indicate that for a single
Gaussian with $\tau_0=3.4$ for the weak line, Eq.~\ref{N} overestimates
$N$ by 4\% in the worst possible case of intermediate
resolution.\footnote{This occurs when the FWHM of a Gaussian
instrumental profile is equal to $1.5b$.}

\subsection{A Cluster of Nonoverlapping Gaussians}\label{sparse_cluster}

We now consider the problem of analyzing several Gaussian profiles that
are blended together by the instrument, but that are actually well
separated from each other in the absence of such blending.  The key to
solving this problem relies on an interesting property of the combined
strengths of saturated lines noted by Jenkins \markcite{1355} (1986). 
He demonstrated that the application of a curve of growth (or doublet
ratio) analysis to a sum of the equivalent widths of an ensemble of
Gaussian profiles, while technically an incorrect procedure, still gives
an answer for the sum of $N$ over all of the contributions that is
usually only slightly below the correct value.  The method works
satisfactorily even when there is a significant dispersion in the
$\tau_0$ and $b$ values for the separate components.  For example, when
the doublet ratio analysis is tried on the two sums of equivalent widths
of population of lines that have an average central optical depth for
the weak line $\langle\tau_0\rangle=4$ but with an rms dispersion of
40\% for the individual $\tau_0$'s and $b$'s, the standard calculation
for the total value of $N$ should, on average, be equal to 94\% of the
true value.  Lower values for either $\langle\tau_0\rangle$ or the
dispersion of $\tau_0$ give results that are closer to the true one. 
The only condition where the analysis breaks down seriously is when the
distribution of $\tau_0$ or $b$ is markedly bimodal (or when there are
so few components that small-number statistical fluctuations may be
important).

From the findings presented in the above paragraph and reasoning given
in \S\ref{single_line}, we can see that a severely underresolved
recording of a sparse cluster of saturated Gaussian profiles can be
treated in the same manner as a single Gaussian.  Once again, we
evaluate the correction factor $C_R(v)$ everywhere, and, through the use
of Eq.~\ref{N}, obtain the column density for the entire group.  This
argument applies even if the cluster of features spans a velocity
interval that is much larger than the instrumental profile.  In this
case, the value of $\tau_a(v)$ at any point represents simply a weighted
sum of the equivalent widths of a bunch of lines on either side of $v$. 
The derived differential value of $N$ at this velocity will reflect,
with reasonable accuracy, the amount of material present in these
components, after their contributions have been multiplied by the same
weight factors.

As with the single Gaussian, an observation that fully resolves the
individual components will yield $C_R(v)=1$ everywhere, with the outcome
that $\tau_a(v)$ needs no upward correction because it is virtually the
same as $\tau(v)$.  Only in the case where the resolving power is
roughly comparable to the line widths do we face the prospect of having
a mild inaccuracy in the answer.

\subsection{A Dense Cluster of Overlapping Gaussian
Profiles}\label{dense_cluster}

Suppose that we now enhance the absorption of the sparse complex of
lines by adding many new components, all of which have the same
population characteristics as before.  If the complex's velocity span
remains constant, we could reach a point where the lines start to
overlap each other.  Going still further, new components could be added
to the point that they pile on top of each other many times over.  We
are now presented with an overall absorption profile that is very
complex and, on average, with a depth $I_0-I_a(v)$ that could now be a
very significant fraction of the continuum level $I_0$.

In real life, if there are no coherent physical processes that can have
a dynamical influence on the radial velocities of the components, one
can expect the small-scale placement of the components' velocity
centroids to be random.  When this is the case, the average optical
depth will grow in direct proportion to the density of components per
unit velocity.  In essence, Beer's law is operating on the average
amount of flux that is permitted to penetrate a random superposition of
absorbers.\footnote{By analogy, one can picture the attenuation of light
passing through a forest.  Averaged over some solid angle of reasonable
size, we expect an exponential attenuation law to apply, even when there
is a mix of large and small trees and, moreover, some of the trees are
opaque while others are translucent.  (However, in contrast to a natural
forest, a commercial tree farm does not obey this principle because
there is some regularity in the locations of the trees.)} This behavior
insures that $\tau_a(v)$ will continue to represent, within the
instrument's passband centered on $v$, a (weighted) sum of equivalent
widths of the components, i.e., the $W_\lambda$'s that would have been
accumulated in the absence of any mutual obstruction of the lines.  The
situation here is therefore no different than that presented in
\S\ref{sparse_cluster}, except that the $\tau_a(v)$'s are no longer much
less than 1.  It follows from the proportionality of line density to
average optical depth that the evolution of a given population of
components from a sparse to a dense cluster has no effect on either
$R(v)$ or the factor $C_R(v)$ that is needed to boost the equivalent
widths (and hence $\tau_a(v)$) to a value that approximately represents
the smoothed $\tau(v)$.

\section{A Demonstration}\label{demo}

\begin{figure}
\plotone{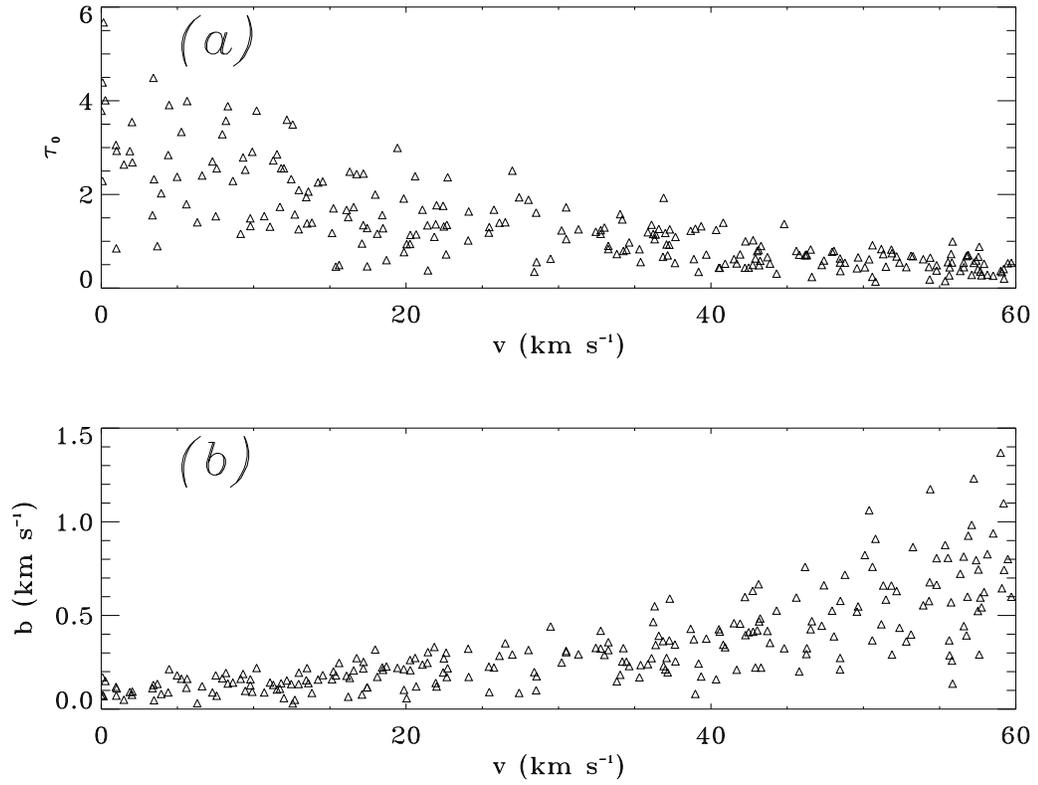}
\caption{Central optical depths $\tau_0$ (panel {\it a\/}) and velocity
dispersions $b$ (panel {\it b\/}) of contributions that made up the
dense line complex demonstration example, as a function of the locations
of the lines' central velocities.\label{population}}
\end{figure}

Now that the basic principles of correcting $\tau_a(v)$ have been
presented, it is appropriate to create some test cases that demonstrate
how the method works.  Two examples will be offered.  The first will
consist of a 60 km~s$^{-1}$ wide complex of mostly saturated Gaussian
lines, all different from each other, whose narrowest members are
distributed sparsely enough to create only occasional overlaps.  The
second example will contain a much denser collection with many more such
lines.  The second example will have a density of components per unit
velocity that is large enough to cause a substantial compounding of
absorptions at most velocities.  These two examples are tailored to
illustrate the principles discussed in \S~\ref{sparse_cluster} and
\S~\ref{dense_cluster}, respectively.
\placefigure{population}

Certain properties of the line complexes were contrived to serve a
useful pedagogical purpose, but at some expense in realism.  For
instance, Fig.~\ref{population} shows the values of central optical
depths and velocity dispersions of components as a function of their
location (velocity) within the dense absorption complex.  (The sparse
complex is just a small subset from the same general population.)  In
both cases, the complexes were constructed such that the average central
optical depths $\langle\tau_0\rangle$ decreased from left to right,
while at the same time the average velocity widths $\langle b\rangle$
increased to make the product $\langle\tau_0\rangle\langle b\rangle$
constant over the entire velocity span.  One could imagine this
absorption complex arising from some fictitious case where a
heterogeneous collection of interstellar clouds had a steady increase in
kinetic temperature as the radial velocities progressed from 0 to
60~km~s$^{-1}$.  Within any small velocity interval, the dispersions of
$\tau_0$ and $b$ were constructed to be equal to 40\% of their mean
values (but with a cutoff at $-2\sigma$ to prevent negative or
inordinately low values).
\placefigure{examplex}
\begin{figure}
\plotone{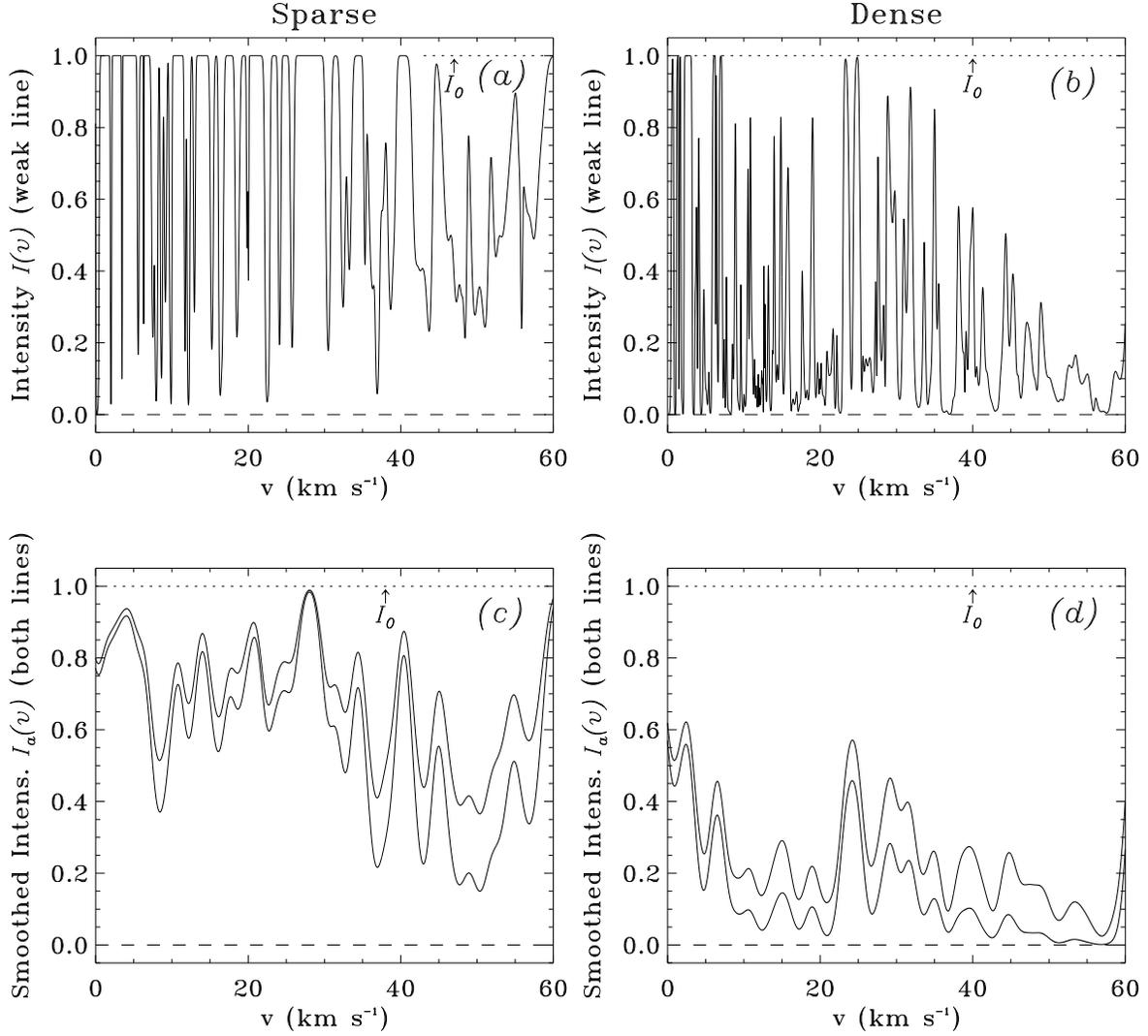}
\caption{Various forms of the two spectral complexes chosen
for the demonstration of corrections of $\tau_a(v)$, plotted against
radial velocity $v$.  The left-hand panels ({\it a}, {\it c}, {\it e\/}
and {\it g\/}) apply to the sparse layout of spectral components, while
the right-hand ones ({\it b}, {\it d}, {\it f\/} and {\it h\/}) depict
the dense complex.   From top to bottom are displayed, ({\it a\/} and
{\it b\/}) the intensities of the weak line before instrumental
smoothing, ({\it c\/} and {\it d\/}) the intensities of both lines after
they have been smoothed by the instrument (upper curve: weak line, lower
curve: strong line), ({\it e\/} and {\it f\/}) the derived values of
$\tau_a(v)$ (lower curve: weak line, upper curve: strong line), and
({\it g\/} and {\it h\/}) the ratios of the two $\tau_a(v)$'s.  The low
values of $R(v)$ on the left-hand side, relative to those on the right,
indicate the need for a stronger correction $C_R(v)$ for the ensembles
of individual components that are deeper and narrower.}\label{examples}
\end{figure}
\begin{figure}
\figurenum{2}
\plotone{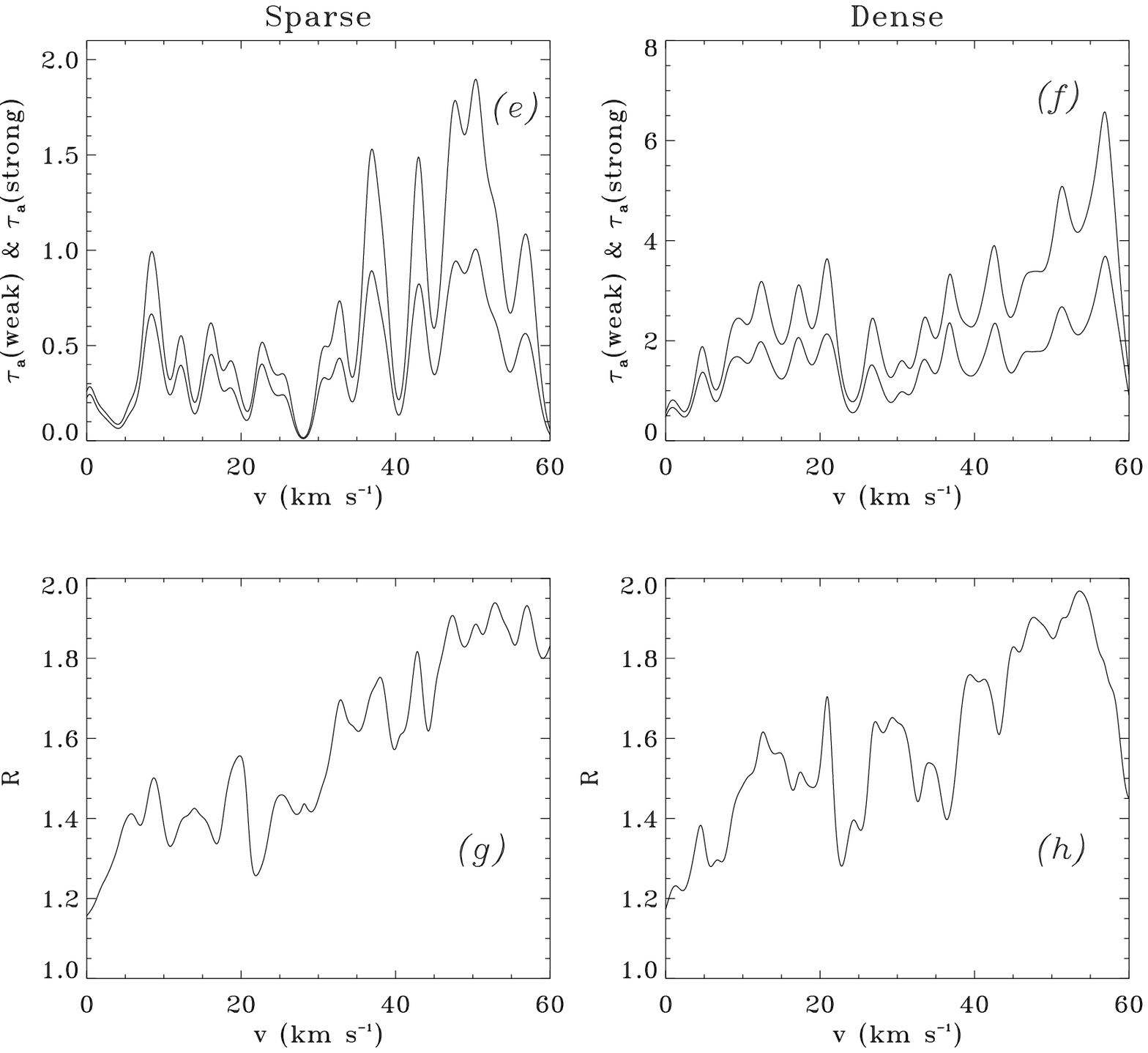}
\caption{continued.}\label{examplex}
\end{figure}
\notetoeditor{If possible, please have the two parts of figure 2 on
facing pages.}
The top two panels ({\it a} and {\it b\/}) in Fig.~\ref{examplex} show
the original, fully resolved intensities of the weak transition for the
sparse and dense ensembles, before they were smoothed by an instrumental
profile that consists of a Gaussian with a FWHM = 2~km~s$^{-1}$. 
Immediately below these panels are the smoothed forms for both the
strong and weak lines.  The intensities are converted to apparent
optical depths and shown in the next row of panels.  In both cases, the
last pair of panels show that the ratios $R(v)$ of the two optical
depths progress in a somewhat irregular fashion from about 1.2 on the
left hand side to about 1.9 on the right.  The fluctuations on top of
the general trend are caused by random changes in the relative mix of
lines with different degrees of saturation.
\placefigure{outcomes}

\begin{figure}
\plotfiddle{ejenkins_fig3.ps}{3truein}{90}{60}{70}{-220}{0}
\caption{Outcomes from the analyses of the sparsely ({\it a\/}) and
densely populated ({\it b\/}) demonstration examples.  Logarithmic
representations of various forms of the the weak line's $\tau$'s are
shown as a function of radial velocity $v$.  In each panel of the
figure, the solid line shows a smoothed version of the real $\tau(v)$
(not known to an observer).  The dotted line is the raw, uncorrected
$\tau_a(v)$.  The dashed line, often so close to the solid line that it
can not be seen, represents the corrected apparent optical depth, 
$C_R(v)\tau_a(v)$.  The curve that encloses the solid shading above the
horizontal line at an ordinate of $-$2.3 illustrates the magnitude of
the error in the uncorrected $\tau_a(v)$, i.e., the function shown is
$\tau(v)_{\rm smoothed}-\tau_a(v) - 2.3$.  Just above this curve is
shown the much smaller error that arises after the $\tau_a(v)$ has been
corrected, i.e., here the plot shows $\tau(v)_{\rm
smoothed}-C_R(v)\tau_a(v) - 1.5$.}\label{outcomes}
\end{figure}

At each velocity $v$, we measure $R(v)$, the ratio of the strong line's
$\tau_a(v)$ to that of the weak one, and derive $C_R(v)$ by solving
Eqs.~\ref{R} and \ref{C} (a streamlined way of directly computing
$C_R(v)$ will be presented in \S~\ref{analytical}).  The results are
shown in Fig.~\ref{outcomes}.  On the left hand side where the
individual components are narrow and saturated, $\tau_a(v)$ (shown by
the dotted line) is significantly below a smoothed version of the true
optical depth $\tau(v)$ (solid line).  Toward the right, the individual
components become less saturated but broader, and this reduces the
disparity.  Note, however, that apart from small-scale random
fluctuations, a running average of the true optical depths (and hence
column density) does not systematically change from left to right.

After multiplying $\tau_a(v)$ by $C_R(v)$, we obtain a corrected optical
depth (dashed line in Fig.~\ref{outcomes}) that is a good approximation
to the smoothed $\tau(v)$ (solid line -- the function that we are
attempting to reconstruct so that we can derive the column density per
unit velocity).  The difference between the two is shown by the
filled-in curve associated with a baseline that is displaced vertically
to a plot $y$ value of $-$1.5.  In the sparse line case, this error is
always less than 0.1 dex, as indicated by the thickness of the black
filling.  For the dense line example, this error occasionally becomes of
order 0.1 dex.  These errors are significantly less than the gross
underestimates for the smoothed $\tau(v)$ that arise from the raw values
of $\tau_a(v)$, as shown by the filled-in curve above the baseline
situated at $-$2.3 in the two panels of the figure.  It is important to
realize, however, that these tests are being performed under conditions
where there is absolutely no noise or systematic errors in defining
either a continuum level or zero intensity baseline.  Under the
circumstances of real observing, one would need to assess the impact of
these errors on the reliability of the results.

\section{A Comparison with the Method of Savage \&
Sembach}\label{comparison}

In the preceding section we have examined the performance of the optical
depth correction technique on a broad expanse of many randomly situated
Gaussian profiles.  It is also useful to test the method on a small
clump of a few such profiles.  Some test examples constructed by Savage
\& Sembach \markcite{110} (1991) are suitable for this purpose.  Their
six cases range in complexity from a single Gaussian (their Case~1) to a
group of 6 components (their Case~6) that overlap and have values of
$\tau_0$ and $b$ that differ by factors of 5 (see their Table~2 for
details).

We also wish to examine how well total column densities derived from the
on-the-spot corrections discussed here compare with the simple global
correction scheme discussed by Savage \& Sembach.  Basically, they
proposed that the total column density (in contrast to the column
density per unit velocity) be compensated according to the disparities
of apparent column densities from the two lines.  It is clear that their
technique would be inappropriate for certain configurations, such as the
two highly contrived cases presented in \S\ref{demo} or certain real
observations [e.g., such as the Fe~II absorption lines in the spectrum
of $\pi$~Sco shown by Joseph \& Jenkins \markcite{1728} (1991)].  For
the demonstration examples in \S\ref{demo}, a correction for the left
hand side would not work for the right hand side where practically no
correction is needed.  However, in some more realistic situations where
just a few saturated and unsaturated components are mixed together, the
story might be different.
\placetable{cases}

\begin{deluxetable}
{
c     
c     
c     
c     
c     
c     
}
\tablecolumns{6}
\tablewidth{400pt}
\tablecaption{Percentage Errors\tablenotemark{a}\label{cases}}
\tablehead{
\colhead{} & \colhead{} & \colhead{Weak Line} &
\multicolumn{3}{c}{Smoothing Profile FWHM (km~s$^{-1}$)}\\
\cline{4-6}\\
\colhead{n\tablenotemark{b}}&\colhead{Case\tablenotemark{c}}
&\colhead{Max. $\tau$}&\colhead{10}&\colhead{20}&\colhead{40}\\
}
\footnotesize
\startdata
2&1&0.4&0.0, $-$0.3&0.3, $-$0.3&0.4, $-$0.2\nl
&2&0.5&0.1, 0.0&0.4, 0.1&0.6, 0.3\nl
&3&0.8&$-$0.1, $-$0.7&0.4, $-$0.8&0.8, $-$0.5\nl
&4&0.9&0.3, $-$0.2&1.1, 0.4&2.0, 1.0\nl
&5&0.8&0.3, $-$0.2&1.2, 0.1&1.7, 0.7\nl
&6&0.4&0.0, 0.9&0.1, $-$0.1&0.4, $-$0.2\nl
3&1&0.8&0.0, $-$0.9&0.4, $-$0.8&1.0, $-$0.5\nl
&2&0.9&0.5, $-$0.1&1.5, 0.4&2.1, 1.0\nl
&3&1.6&$-$0.5, $-$2.2&0.1, $-$1.8&0.5, $-$1.1\nl
&4&1.8&1.4, $-$0.1&4.1, 2.4&6.2, 4.6\nl
&5&1.7&1.3, $-$0.5&4.1, 1.2&5.6, 3.1\nl
&6&0.8&0.1, $-$0.2&0.3, $-$0.4&0.8, $-$0.4\nl
4&1&1.7&$-$0.7, $-$2.4&$-$0.1, $-$1.7&0.4, $-$0.9\nl
&2&1.8&2.4, 0.1&5.0, 2.0&6.3, 3.8\nl
&3&3.1&$-$1.6, $-$3.3&$-$2.2, $-$1.3&$-$3.6, $-$0.7\nl
&4&3.5&6.6, 3.1&13.9, 11.2&18.0, 16.2\nl
&5&3.4&6.4, 1.1&12.6, 11.2&15.3, 10.5\nl
&6&1.5&0.7, $-$0.4&0.8, $-$0.9&2.3, 0.4\nl
5&1&3.4&$-$1.0, $-$2.1&$-$2.3, 0.0&$-$4.5, 0.3\nl
&2&3.6&8.7, 3.2&13.6, 8.1&15.0, 10.7\nl
&3&6.3&2.5, 4.0&0.2, 7.0&$-$3.8, 4.6\nl
&4&7.1&20.9, 16.1&31.4, 28.9&36.2, 34.8\nl
&5&6.7&18.8, 9.8&26.9, 17.9&29.1, 22.7\nl
&6&3.0&2.7, 0.1&3.7, 1.0&9.7, 8.1\nl
\enddata
\tablenotetext{a}{Double entries separated by a comma show the error
using the method of Savage \& Sembach \protect\markcite{110} (1991)
(first number) and the on-the-spot corrections discussed here (second
number).  Percentages represent $100\times[N_{\rm true}-N_{\rm
computed}]/N_{\rm true}$, where $N$ is the integrated column density for
the entire group of components in each case.}
\tablenotetext{b}{Exponent of 2 in a multiplicative scale factor for all
of the components.  See Eq.~12 of Savage \& Sembach.}
\tablenotetext{c}{Identification of the particular component group. See
Table~2 of Savage \& Sembach.}
\end{deluxetable}

In the limit that the resolving power is extremely poor, the two methods
should give identical results (and, conversely, at very high resolution
no correction is needed).  Differences should be apparent only at
intermediate resolutions.  Table~\ref{cases} shows the percentage
errors, $100\times[N_{\rm true}-N_{\rm computed}]/N_{\rm true}$, for
both methods, applied to the 6 cases of Savage \& Sembach with various
scaling factors for the optical depths proportional to $2^n$.  At a
resolution of 10~km~s$^{-1}$ the two methods perform about equally,
i.e., in 14 of the 24 examples the on-the-spot corrections have a
smaller error than global correction method of Savage \& Sembach, and
for the remaining 10 the converse is true.  At 20~km~s$^{-1}$ the
on-the-spot technique does about twice as well (15 wins vs. 7 losses,
with 2 draws), and at 40~km~s$^{-1}$ the on-the-spot technique is
superior in 21 cases and not so in 3.  

One may notice that the column density correction factors listed by
Savage \& Sembach (in their Table~4) for adjusting an isolated Gaussian
component differ slightly from the $C_R$ that comes from solving
Eqs.~\ref{R} and \ref{C}.  For $\log(R/2)$ $[\equiv \log N^{n-1}_a -
\log N^n_a$ in the notation of Savage \& Sembach] $< 0.18$, the two are
very close to each other.  However if $\log(R/2) = 0.22$, the logarithm
of the correction factor for equivalent widths is 0.418, while the
corresponding value listed by Savage \& Sembach is 0.453.  It is
possible that the difference is explained by the fact that their
correction factors are adjusted slightly to give better performance at
intermediate resolving powers.  This conclusion seems to be supported by
the results shown in Table~\ref{cases}.  The method of Savage \& Sembach
works very well for the pure Gaussian profile (Case 1) and an assemblage
of profiles that look very similar to a single Gaussian (Case 3).

\section{An Example using Real Observations}\label{real_obs}

As a final demonstration, we explore how well the corrected apparent
optical depths reproduce the results of a component model derived from
real observations.  Spitzer \& Fitzpatrick \markcite{2462} (1993)
identified 9 velocity components of S~II toward HD~93521 using the
absorption profiles from the transitions at 1251\AA\ ($\log f\lambda =
0.837$) and 1254\AA\ ($\log f\lambda = 1.136$), along with some
supporting information from the absorption features of other species. 
One can see from their Fig.~2$a$ that these profiles are moderately
saturated.  However, at the resolution of the GHRS echelle spectrograph,
it appears that the lines are fully resolved because $\tau_a(v)$ of the
weaker line is, to within observational errors, always equal to half
that of the stronger line at all velocities.  Thus, no correction is
needed, and raw $\tau_a(v)$ values from either line are appropriate for
deriving the column density per unit velocity.\footnote{This good
resolution of the velocity structure by the GHRS echelle spectrograph is
not generally achieved when the spectra of stars behind cold clouds are
recorded.  The line of sight to the high latitude star HD~93521 seems to
penetrate only warm material that has components with $b$ values that
are generally about equal to or greater than the width of the
instrumental profile.}

\begin{deluxetable}
{
r     
r     
r     
r     
r     
r     
r     
r     
}
\tablecolumns{8}
\tablewidth{0pt}
\tablecaption{Coefficients for Eq.~\protect\ref{logC}\label{coef}}
\tablehead{
\colhead{Ratio} & \colhead{} & \colhead{} & \colhead{} &
\colhead{} & \colhead{} & \colhead{Max.} & \colhead{Min.}\\
\colhead{of $f\lambda$} & \colhead{$a_0$}& \colhead{$a_1$} 
& \colhead{$a_2$}& \colhead{$a_3$}& \colhead{$a_4$} & 
\colhead{Residual\tablenotemark{a}} & \colhead{$R$}\\
}
\startdata
$2^{0.50}$&0.00096&0.29122&$-$0.01695&$-$0.03787&0.02525&$-$0.00096&1.07\nl
$2^{0.75}$&0.00106&0.24182&0.01361&$-$0.04585&0.02407&$-$0.00106&1.10\nl
$2^{1.00}$&0.00092&0.20322&0.02549&$-$0.04102&0.02002&0.00124&1.14\nl
$2^{1.50}$&0.00049&0.14628&0.02569&$-$0.02118&0.01103&$-$0.00049&1.25\nl
$2^{2.00}$&0.00022&0.10528&0.02029&$-$0.00755&0.00567&0.00110&1.40\nl
$2^{2.50}$&0.00003&0.07625&0.01296&0.00128&0.00252&$-$0.00013&1.70\nl
$2^{3.00}$&0.00001&0.05413&0.01043&0.00249&0.00183&0.00022&1.90\nl
\enddata
\tablenotetext{a}{Defined as $\log (C_R)_{\rm exact}-\log (C_R)_{\rm
Eq.~\protect\ref{logC}}$ over the range Min($R$) $<R<$ Ratio of
$f\lambda$.}
\end{deluxetable}

\begin{figure}[t!]
\epsscale{0.5}
\plotone{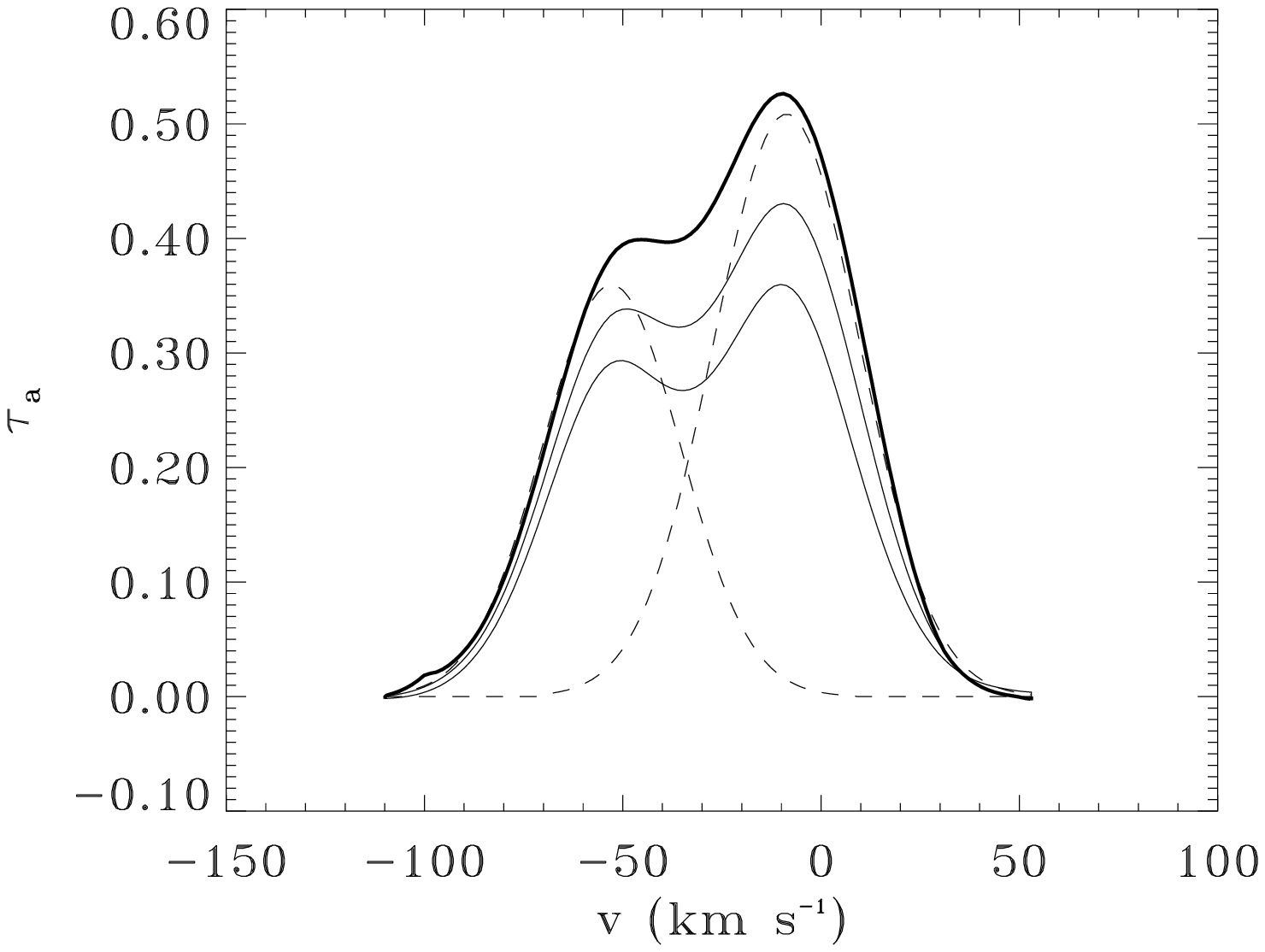}
\caption{Plots of the apparent optical depths of the S~II absorption
lines in the spectrum of HD 93521, evaluated from the data of Spitzer \&
Fitzpatrick \protect\markcite{2462} (1993) after being convolved with a
Gaussian profile with a FWHM = $36.7~{\rm km~s}^{-1}$ (i.e., $\sigma$ =
10 data points) to simulate the effects of smoothing by a low resolution
spectrograph.  The upper thin curve represents $\tau_a(v)$ for the weak
transition at 1251\AA, while the lower thin curve is $\tau_a(v)/2$ for
the strong one at 1254\AA.  The thick curve represents $C_R(v)\tau_a(v)$
for the weak line, which can be approximately broken down as a sum of
two Gaussian components (dashed curves).}\label{sii}
\end{figure}

To make the demonstration nontrivial and create a disparity in the
results for the two transitions, we can smooth the profiles so that they
are significantly under-resolved.  This smoothing reconstructs how the
absorption features would appear if they were observed with a
low-resolution spectrograph.  The two thin curves in Fig.~\ref{sii} show
the $\tau_a(v)$ relationships for these two S~II transitions whose
$f\lambda$'s differ by a factor of 2.  The $\tau_a(v)$ for the stronger
transition was divided by 2 before it was plotted, in order to show how
much its implied column density per unit velocity differs from that of
the weaker transition.
\placefigure{sii}

The heavy line in Fig.~\ref{sii} shows the corrected $\tau_a(v)$ for the
weaker line, i.e., the appearance of the line after it has been
multiplied by $C_R(v)$.  $C_R(v)$ varies from 1.0 to about 1.25 over
most of the relevant velocity range of the profiles.  An application of
Eq.~\ref{N} yields $N=2.14\times 10^{15} {\rm cm}^{-2}$.  By comparison,
the detailed model for the S~II components given by Spitzer \&
Fitzpatrick gives a total column density of $2.10\times 10^{15} {\rm
cm}^{-2}$.  If we decompose our corrected $\tau_a(v)$ into two Gaussian
components (dashed lines in Fig.~\ref{sii}), we obtain $N=8.61\times
10^{14} {\rm cm}^{-2}$ for the left-hand one and $N=1.29\times 10^{15}
{\rm cm}^{-2}$ for the one on the right.  These values are close to the
sums of column densities in two distinct bunches of components in the
data of Spitzer \& Fitzpatrick.  Their components 1$-$4 had $8.83\times
10^{14} {\rm cm}^{-2}$, and 5$-$9 had $1.22\times 10^{15} {\rm
cm}^{-2}$).  Evidently, within the limitations of what can be seen at
the low resolution, the profile representing $C_R(v)\tau_a(v)$ is
consistent with detailed component fits performed by Spitzer \&
Fitzpatrick on their high-resolution spectrum.

\section{Analytical Approximations for $C_R$}\label{analytical}

The solutions to Eqs.~\ref{R} and \ref{C} are somewhat awkward to
compute in a data reduction program.  Thus, as a convenience for
observers who wish to undertake an analysis that invokes the corrections
for $\tau_a(v)$, some simple analytical approximations for $C_R$ will be
presented.  Since these formulae can be differentiated, they are also
useful for evaluating how both random and systematic errors respond to
the transformations.

It turns out that the relationship between a variable
\begin{equation}\label{y}
y=\ln(1+\log C_R)
\end{equation}
and
\begin{equation}\label{x}
x=\ln\left[ {(f\lambda)_{\rm strong}\over (f\lambda)_{\rm
weak}}-1\right] - \ln (R-1)
\end{equation}
is remarkably close to linear.  The small departures from linearity can
be appropriately handled by a low order polynomial whose coefficients
may be evaluated by a least squares fit to the exact calculations.  In
the practice of reducing real data, we are now free to disregard
Eqs.~\ref{R} and \ref{C} and, instead, use the much simpler explicit
equation
\begin{equation}\label{logC}
\log C_R=\exp(a_0 + a_1x + a_2x^2 + a_3x^3 + a_4x^4) - 1
\end{equation}
to obtain (the logarithm of) the correction factor for the weak line's
$\tau_a(v)$.

Table~\ref{coef} lists the coefficients $a_0$ through $a_4$ for 6
different values for the ratios of $f\lambda$ for the two lines (the
third row corresponds to the doublet ratios that were considered in the
previous examples).  The coefficients other than $a_1$ correct for the
small departures from linearity.  Note that for a good general fit,
$a_0\neq 0$ even though the intercept at $x=0$ (corresponding to no
saturation at all) should be 0.  The next to the last column in the
table shows the largest deviations for $\log C_R$ between the results of
Eq.~\ref{logC} and the exact solutions to Eqs.~\ref{R} and \ref{C} down
to a minimum value of $R$ shown in the final column.  All of the
coefficients vary with $f\lambda$ ratios in a smooth manner.  Thus, one
can obtain coefficients for arbitrary ratios between those listed in the
table by interpolation.  
\placetable{coef}

\section{Summary: Some Cautions and a Recipe}\label{recipe}

The basic principles developed in \S\ref{concepts} and the
demonstrations in \S\ref{demo} and \S\ref{real_obs} show that it is
possible, even under fairly harsh circumstances of unresolved
saturation, to derive good measurements of the column densities of
absorbing substances as a function of velocity if two or more lines with
different transition probabilities are observed.  The analysis method
proposed in this paper has a special virtue, in that it avoids the
requirement for model building: one is not forced to try to reconstruct
exactly what profiles should have looked like before they were smoothed
by the instrument.

Obviously, how well the method works depends on a number of experimental
conditions, such as ratio of $f\lambda$ of two lines, the signal-to-nose
ratio of the spectrum, how well various kinds of systematic error are
controlled, and the accuracy in the match of the two velocity scales. 
It is not easy to give guidelines here on these issues, since there are
so many possibilities.  A formal evaluation of how the  errors can
propagate through the mathematical transformations can give some
guidance.  This sort of error analysis combined with common sense are
probably the best tools for judging the reliability of the final
results.

One must be wary of the trap where a disproportionately large fraction
of the gas is located in very narrow features that are fully saturated
in both lines, but where most of the absorption is caused by
high-velocity fluff that is unsaturated.  This condition is especially
likely to arise when there is a bimodal distribution in the widths and
strengths of the components in the ensemble.  One should be alert for
physical circumstances that could produce this state, such as lines of
sight that have the right mixture of contributions from cold H~I and
much hotter H~II regions, cases where both quiescent and
shock-accelerated gases are present, or, when one is looking through
entire galaxies, there is a mixture of disk and halo gases.  That being
said, it is reassuring to see from the demonstration in \S\ref{demo}
that the analysis is remarkably tolerant to the existence of components
that span a wide range of properties (as long as they are not too
bizarre).  Even a power law for the distribution of $\tau_0$ is
acceptable: see Jenkins \markcite{1355} (1986) for details.

To summarize how the correction method works, we review in the form of a
recipe how one would apply it in practice:
\begin{enumerate}
\item For a given absorber, obtain spectra of two lines that have values
of $f\lambda$ that differ by a large enough factor to show when
saturation might be taking place.
\item On the basis of what the lines look like, coupled with any
independent evidence (e.g. higher quality observations of other species)
or a general knowledge of the physical situation, decide that the
pitfall discussed in the above paragraph does not apply.  If it could,
do not proceed further.  Also, the analysis should not be undertaken if
$\tau_a(v)$ is very large (the threshold for rejection should depend on
how accurately the intensities are measured).
\item For each line, measure $I_0/I_a(v)$ and evaluate its natural
logarithm to obtain $\tau_a(v)$.
\item At every velocity $v$, use Eq.~\ref{x} to evaluate the quantity
$x(v)$ from measurements of $R(v)$, the ratio of the strong line's
$\tau_a(v)$ to that of the weak line, along with the two lines'
$f\lambda$'s.
\item From $(f\lambda)_{\rm strong}/(f\lambda)_{\rm weak}$, determine
appropriate values for the coefficients $a_0$ through $a_4$ from
Table~\ref{coef}.  If necessary, intermediate values can be found by
interpolation.
\item For every $v$, evaluate the correction $C_R(v)$ by the use of
Eq.~\ref{logC} and the coefficients derived in the preceding step.
\item Derive the column density of the absorber per unit velocity by
taking the product of the weak line's $\tau_a(v)$ and its enhancement
factor $C_R(v)$, and multiplying it by the constants in front of the
integral in Eq.~\ref{N}.  Numerically, the latter equals $3.767\times
10^{14}/f\lambda$, if $\lambda$ is expressed in \AA\ and the column
density per unit velocity has the units ${\rm cm}^{-2}({\rm
km~s}^{-1})^{-1}$.
\end{enumerate}
\acknowledgments
This research was supported by NASA grant NAG5-616 and grants
GO-2403.02-87A and GO-2344.01-87A from the Space Telescope Science
Institute.  The inspiration for the development of the correction method
described here came from data recorded by the Interstellar Medium
Absorption Profile Spectrograph (IMAPS), an ultraviolet spectrograph
that flew on the Astrospas orbital mission in September 1993.  In spite
of the high wavelength resolving power of this instrument, it was clear
that some profiles needed to have an adjustment of their $\tau_a$'s
because they led to discrepancies in the inferred column densities for
lines of different strength.  The author is indebted to C. Joseph, B.
Savage, K. Sembach and L. Spitzer for their useful comments on an early
draft of this paper.

\end{document}